\documentclass[aps,prl,twocolumn,superscriptaddress]{revtex4-2}
\usepackage[utf8]{inputenc}
\usepackage{dsfont}
\usepackage{amsfonts}
\usepackage{amsmath}
\allowdisplaybreaks[4]        
\usepackage{amssymb}
\usepackage{euscript}     
\usepackage{braket}
\usepackage{starfont}
\usepackage{color,soul}         
\usepackage{tensor}        
\usepackage{amsthm}
\usepackage{graphicx}
\usepackage{slashed}
\usepackage{leftidx}
\usepackage{subfigure}
\usepackage{bbm}
\definecolor{outerspace}{rgb}{0.25, 0.29, 0.3}
\definecolor{scarlet}{rgb}{1.0, 0.13, 0.0}
\usepackage[header,title,page,titletoc]{appendix}  
\definecolor{princetonorange}{rgb}{1.0, 0.56, 0.0}
\definecolor{WildStrawberry}{rgb}{1.0, 0.26, 0.64}
\definecolor{rossocorsa}{rgb}{0.83, 0.0, 0.0}
\definecolor{navyblue}{rgb}{0.0, 0.0, 0.5}
\usepackage{float}
\usepackage[paper=letterpaper,margin=1in]{geometry}

\usepackage{mathtools}

\usepackage{wasysym}
\usepackage{hyperref}
\hypersetup{
    colorlinks,
    citecolor=Peach,
    filecolor=Peach,
    linkcolor=MidnightBlue,
    urlcolor=MidnightBlue
}

\DeclareMathAlphabet{\pazocal}{OMS}{zplm}{m}{n}
\newcommand{\req}[1]{(\ref{#1})} 

\newcommand{\bea}{\begin{eqnarray}}

\newcommand{\eea}{\end{eqnarray}}
\newcommand{\ba}{\begin{eqnarray}}
\newcommand{\ea}{\end{eqnarray}}

\newcommand{\be}{\begin{equation}}
\newcommand{\ee}{\end{equation} }
\newcommand{\beqa}{\begin{eqnarray}}
\newcommand{\eeqa}{\end{eqnarray}}
\newcommand{\beqar}{\begin{eqnarray*}}
\newcommand{\eeqar}{\end{eqnarray*}}

\renewcommand{\req}[1]{(\ref{#1})}

\newcommand{\ie}{{\it i.e.,}\ }

\newcommand{\D}{\mathcal{D}}
\newcommand{\E}{\mathcal{E}}

\makeatletter
\newcommand{\dal}{\mathop{\mathpalette\dal@\relax}}
\newcommand{\dal@}[2]{%
  \begingroup
  \sbox\z@{$\m@th#1\square$}%
  \dimen0=\fontdimen8
    \ifx#1\displaystyle\textfont\else
    \ifx#1\textstyle\textfont\else
    \ifx#1\scriptstyle\scriptfont\else
    \scriptscriptfont\fi\fi\fi3
  \makebox[\wd\z@]{%
    \hbox to \ht\z@{%
      \vrule width \dimen0
      \kern-\dimen0
      \vbox to \ht\z@{
        \hrule height \dimen0 width \ht\z@
        \vss
        \hrule height 2\dimen0
      }%
      \kern-2.5\dimen0
      \vrule width 2.5\dimen0
    }%
  }%
  \endgroup
}
\makeatother

\usepackage[dvipsnames]{xcolor}
\usepackage{wasysym}
\usepackage{hyperref}

\hypersetup{
    colorlinks=true,
    citecolor=Peach,
    filecolor=Peach,
    linkcolor=MidnightBlue,
    urlcolor=MidnightBlue
}

\begin{document}

\title{Regular Black Holes in Nonlocal Quasitopological Gravity}
\author{Pablo Bueno}
\email{pablobueno@ub.edu}
\affiliation{Departament de F\'isica Qu\`antica i Astrof\'isica, Institut de Ci\`encies del Cosmos\\
 Universitat de Barcelona, Mart\'i i Franqu\`es 1, E-08028 Barcelona, Spain }

\author{{ Pablo A. Cano}}
\email{pablo.cano@um.es}
\affiliation{Departamento de Física, Universidad de Murcia, Campus de Espinardo, 30100 Murcia, Spain }

\author{ Robie A. Hennigar}
\email{robie.a.hennigar@durham.ac.uk}
\affiliation{Centre for Particle Theory, Department of Mathematical Sciences, Durham University, Durham DH1 3LE, U.K.}

\author{\'Angel J. Murcia}
\email{angelmurcia@icc.ub.edu}
\affiliation{Departament de F\'isica Qu\`antica i Astrof\'isica, Institut de Ci\`encies del Cosmos\\
 Universitat de Barcelona, Mart\'i i Franqu\`es 1, E-08028 Barcelona, Spain }


\begin{abstract}


\noindent 
 We present infinite-derivative completions of Quasitopological gravities that are ghost-free, avoid strong coupling instabilities and admit exact, spherically symmetric vacuum regular-black-hole solutions satisfying a perturbative Birkhoff theorem.


\end{abstract}
\maketitle

{\bf Introduction.} Resolving the curvature singularities of general relativity (GR) remains a central problem in theoretical physics~\cite{Penrose:1964wq}.  Broadly speaking, there are two possible ways this might happen. Singularities may signal the breakdown of spacetime itself, requiring a description in terms of more fundamental quantum degrees of freedom. Alternatively, they may be resolved at an effective classical level through corrections to GR induced by quantum gravity. While the former possibility is difficult to address without a complete theory of quantum gravity, the latter is amenable to bottom-up analyses.



Nonlocal gravity provides one arena in which to pursue this idea~\cite{Biswas:2005qr, Biswas:2011ar,Modesto:2011kw,Tomboulis:1997gg}. Nonlocality can  smear localized sources over finite regions, and thereby regularize fields that would be singular in GR. It also offers a way to avoid the Ostrogradski ghosts associated with theories containing finitely many higher derivatives and can yield renormalizable or finite quantum theories~\cite{Modesto:2011kw}. However, much of the evidence for singularity resolution comes from weak-field studies, while the full nonlocal equations are difficult to control and their solutions remain poorly understood~\cite{Buoninfante:2022ild}. A more serious issue is that, in many nonlocal models, all vacuum solutions of GR remain solutions, including singular ones~\cite{Li:2015bqa,Calcagni:2017sov}. This is problematic because, in GR, spacetime itself can collapse to form black holes and singularities even in the complete absence of matter~\cite{Beig:1991dh,Abrahams:1992ib, Bizon:2005cp, Christodoulou:2008nj}. Thus, any viable mechanism should be capable of resolving singularities already present in vacuum gravity.

A different route to singularity resolution is to make curvature itself bounded~\cite{PismaZhETF.36.214}. Recent work has shown that such a limiting curvature can emerge dynamically from infinite towers of higher-curvature corrections to the Einstein-Hilbert action~\cite{Bueno:2024dgm}. The most concrete implementation of this is provided by quasitopological (QT) gravity~\cite{Oliva:2010eb, Quasi, Dehghani:2011vu, Cisterna:2017umf, Ahmed:2017jod,Bueno:2019ycr, Bueno:2022res, Moreno:2023rfl,Moreno:2023arp}: a class of higher-curvature theories, defined in dimensions $D\geq 5$, whose polynomial curvature invariants provide a basis for the gravitational effective action~\cite{Bueno:2019ltp}, yield second-order equations in spherical symmetry, and satisfy a Birkhoff theorem~\cite{Bueno:2025qjk}. Under mild assumptions on the couplings, the unique spherically symmetric solutions of these theories are regular black holes (RBHs).\footnote{For early phenomenological work on RBHs see, for example,~\cite{Sakharov:1966aja,1968qtr..conf...87B,1981NCimL......161G,Poisson:1988wc,Dymnikova:1992ux,Hayward:2005gi}; for more recent work see~\cite{Ayon-Beato:1998hmi,Bronnikov:2000vy,Kunstatter:2015vxa,Carballo-Rubio:2018pmi, Carballo-Rubio:2022kad, Barenboim:2025ckx,Carballo-Rubio:2025fnc,Carballo-Rubio:2026gwg}.} This provides a concrete mechanism for singularity resolution in vacuum.


The well-behaved spherical dynamics of QT gravity makes it a useful framework for addressing longstanding questions about RBHs~\cite{Konoplya:2024hfg, DiFilippo:2024mwm, Konoplya:2024kih, Ma:2024olw, Ditta:2024iky, Frolov:2024hhe, Bueno:2024eig, Bueno:2024zsx, Bueno:2025zaj, Hennigar:2025yqm, Bueno:2025tli,Aguayo:2025xfi,Fernandes:2025fnz,Cisterna:2025vxk, Frolov:2025ddw,Arbelaez:2025gwj,Hao:2025utc,Fernandes:2025mic,Frolov:2026rcm, Bueno:2026dln, Arbelaez:2026eaz,DiFilippo:2026jpv,Dubinsky:2026gcj,Borissova:2026rbi,PinedoSoto:2026hfm,Tsuda:2026xjc,Sueto:2026epz}.  However, the approach has important limitations. First, while the equations of motion on general backgrounds are of fourth order, the equations on spherical symmetry and the linearized equations around maximally symmetric backgrounds degenerate to second order, potentially leading to a strong-coupling problem \cite{BeltranJimenez:2020lee,Delhom:2022vae}. Second, the fourth-order character of the equations raises the possibility of Ostrogradski instabilities.


In this Letter, we combine these two approaches. We construct nonlocal extensions of QT gravity which retain the exact vacuum RBHs of the QT theories while using nonlocal form factors to cure their structural problems. The resulting theories remove the strong-coupling problem and, for suitable form factors, introduce no additional ghostlike modes.
We further show that the RBHs admit no nontrivial continuous deformations within the nonlocal theory, establishing a perturbative version of Birkhoff’s theorem. These models therefore unite two mechanisms for singularity resolution that have been hitherto realized separately: the smearing of localized matter sources familiar from nonlocal gravity and the limiting-curvature mechanism responsible for vacuum RBHs in QT gravity.




{\bf Quasitopological Gravity.}
Consider a $D$-dimensional gravitational theory of the form
\begin{equation}\label{QTaction}
S_{\rm QT}= \int \frac{\mathrm{d}^Dx \sqrt{|g|}}{16\pi G} \mathcal{L}_{\rm QT}\,, \quad \mathcal{L}_{\rm QT}=R+\sum_{n=2}^{n_{\rm max}} \alpha_n \mathcal{Z}_n \, .
\end{equation}
Here, the usual Einstein-Hilbert action is supplemented by a tower of $n_{\rm max}-1$ higher-curvature terms. Each density $\mathcal{Z}_n$ is built from certain linear combinations of various contractions of $n$ Riemann tensors and the $\alpha_n$ are free couplings with dimensions of length$^{2(n-1)}$. 
The equations of motion can be written as
\cite{Padmanabhan:2011ex}
\begin{align}\label{eomQT}
\E_{ab}&\equiv P_{acde} R_{b}{}^{cde}-\frac{1}{2}\mathcal{L}_{\rm QT} g_{ab}+2\nabla^c \nabla^d P_{acbd} \\
&=G_{ab}+\sum_{n=2}^{n_{\rm max}}\E_{ab}^{(n)}=0\,, \quad P^{abcd}\equiv \frac{\partial \mathcal{L}_{\rm QT}}{\partial R_{abcd}}\, ,
\end{align}
where $G_{ab}$ is the Einstein tensor and $\E_{ab}^{(n)}$ is the equation of motion tensor of the $n$-th density. The defining property of QT densities is that the term $\nabla^c \nabla^d P_{acbd}$, responsible for the presence of fourth-order derivatives, vanishes identically for general spherically symmetric (SS) configurations  \cite{Bueno:2022res,Bueno:2025qjk},  
\begin{equation}\label{qtdef}
 \mathcal{L}_{\rm QT} \quad \Leftrightarrow \quad  \left.  \nabla^c \nabla^d P_{acbd}\right|_{\rm SS}=0\, .
\end{equation}
As a result, QT gravities  represent generalizations of Lovelock gravities \cite{Lovelock1,Lovelock2} which, remarkably, may be constructed at arbitrarily high order \cite{Bueno:2019ycr,Bueno:2022res,Moreno:2023rfl} for any $D \geq 5$ (see \emph{e.g.}~Section 2 of \cite{Bueno:2025gjg} for explicit expressions).\footnote{In $D=4$ there also exist QT densities of arbitrarily high order which possess RBH solutions satisfying the same equations as in the  $D\geq 5$ cases presented in the text. However, this requires considering non-polynomial densities in the gravitational action \cite{Colleaux:2017ibe,Bueno:2025zaj,Borissova:2026wmn,Borissova:2026krh}. }
The fact that QT theories possess second-order equations for SS spacetimes
implies that the dynamical evolution is under control and free of ghosts in those cases.\footnote{In fact, the full SS sector of general QT theories can be shown to be equivalent to a  certain class of two-dimensional Horndeski theories \cite{Bueno:2024zsx}.} 
Indeed, consider a general SS ansatz,
\begin{equation}
    {\rm d}s^2=-N(t,r)^2f(t,r) {\rm d}t^2+\frac{{\rm d} r^2}{f(t,r)}+r^2{\rm d}\Omega_{(D-2)}^2 \, .
\end{equation}
In vacuum, the full non-linear equations \req{eomQT} get dramatically simplified and are given by \cite{Bueno:2024zsx}
\begin{equation}\label{fNh}
    \partial_t f=0\, , \quad \partial_r N=0\, , \quad \frac{\partial}{\partial r}\left[r^{D-1}h(\psi) \right]=0\, ,
\end{equation}
where we defined the characteristic polynomial
\begin{equation}
    h(\psi)\equiv \psi+\sum_{n=2}^{n_{\rm max}} \alpha_n\psi^n\, , \quad \psi \equiv \frac{1-f}{r^2}\, .
\end{equation}
The first two equations in  \req{fNh} enforce that the most general SS solution of a QT theory is static and characterized by a single function, $f(r)$. This function is a solution of the third equation, namely, $ h(\psi)=m/ r^{D-1}$,
where $m$ is an integration constant proportional to the mass of the solution. When the series of QT densities is truncated at some finite $n_{\rm max}$,
the curvature invariants remain divergent at $r=0$. However, if we allow for infinitely many terms, the interior develops a de Sitter (dS) core and the singularity is completely resolved. This occurs provided $h(\psi)$ satisfies mild qualitative conditions~\cite{Bueno:2024dgm,Fernandes:2025fnz, Bueno:2026dln}.

It is convenient---but not necessary---to assume that the scale of all the $\alpha_n$ is the same. As a paradigmatic example, let us choose $\alpha_n=\alpha^{n-1}$ for certain  $\alpha$ with dimensions of length$^2$ \cite{Bueno:2024dgm}. In that case, we recover the Hayward black hole as a vacuum solution of the corresponding QT theory \cite{Hayward:2005gi}
\begin{equation}\label{hayw}
    f(r)=1-\frac{\mathsf{M} r^2}{r^{D-1}+\alpha \mathsf{M}}\, ,
\end{equation}
where $\mathsf{M}$ is proportional to the ADM mass.
This approaches the Schwarzschild spacetime at long distances, but it contains a regular dS core in its interior: $f(r) =  1-r^2/\alpha+\dots $ as $r\rightarrow 0$. 

Other choices of $\alpha_n$ yield different models, but this example captures the general structure. For each choice of couplings, the corresponding RBHs are the unique 
SS solutions of the QT theories---\ie a Birkhoff theorem holds \cite{Bueno:2025qjk}. Depending on the relative values of $m$ and $\alpha$, the solutions describe globally regular horizonless spacetimes or RBHs with one or two horizons. Remarkably, these RBHs are dynamically formed from the gravitational collapse of matter in simple SS models  \cite{Bueno:2024zsx,Bueno:2024eig,Bueno:2025zaj}.

An important consequence of \req{qtdef} is that both the equations on spherical symmetry and the linearized equations of QT theories around maximally symmetric backgrounds are second order \cite{Quasi,Bueno:2017sui}. Thus, on those backgrounds, QT theories propagate only a massless graviton and are ghost-free.
While this may seem like an attractive feature, it is also the origin of a strong coupling instability.
This arises because the derivative order is reduced on spherical symmetry and maximally symmetric backgrounds, whereas less symmetric backgrounds have fourth-order linearized equations~\cite{BeltranJimenez:2020lee,Delhom:2022vae}.
In mathematical terms, the principal symbol of the equations of motion is pathological because it \emph{depends strongly} on the background. In particular, the principal symbol vanishes on these specific backgrounds, making them a singular point of the equations. 

This raises a natural question: can QT theories be modified so as to cure these pathologies while preserving a healthy SS sector? In the following section, we show that they can.

{\bf Nonlocal Quasitopological Gravity.} To tackle the problem, we add terms involving higher derivatives of the metric to avoid the degeneracy of the equations on SS backgrounds and the linearized equations on maximally symmetric backgrounds.
A key element of our proposal involves the following tensor,
\begin{align}\label{Ehat}
\hat{\E}_{ab}&\equiv P_{acde} R_{b}{}^{cde}-\frac{1}{2}\mathcal{L}_{\rm QT} g_{ab}\, ,
\end{align}
which equals the equations of motion of QT gravity with the $\nabla^c \nabla^d P_{acbd}$ term omitted. We then consider the modified theory
\begin{equation}\label{NLQTv2}
\begin{aligned}
S_{\rm NLQT}=\int \frac{\mathrm{d}^Dx \sqrt{|g|} }{16\pi G}&\left[\mathcal{L}_{\rm QT}-\frac{1}{2}\hat{\E}^{ab} \mathcal{F}_{ab}{}{}^{cd} \hat{\E}_{cd}\right]\, ,
\end{aligned}
\end{equation}
where $\mathcal{F}_{ab}{}{}^{cd}$ is, for the moment, an unspecified covariant operator containing potentially an infinite number of covariant derivatives and the  ``NLQT'' label stands for ``Nonlocal Quasitopological'' gravity. By construction, this theory contains all the solutions of the QT theory satisfying $\hat{\E}_{ab}=0$---in particular, the SS solutions---since the extra piece is quadratic in this tensor. 

If $\mathcal{F}_{ab}{}{}^{cd}$ is truncated to contain at most $2N$ covariant derivatives, the full equations of motion of this theory are of order $2N+4$---this is because $\hat{\E}_{ab}$ contains at most second-order derivatives of the metric. This is also the order of the linearized equations around generic backgrounds for this theory, including maximally symmetric and SS ones. Therefore, the theory avoids strong-coupling problems caused by the reduction of the order of the equations motion. 
Had the theory been built using $\E_{ab}$ instead of $\hat \E_{ab}$, its equations would have been of order $2N+8$ but the linearized equations on maximally symmetric backgrounds (as well as the equations in SS) would have still been of order $2N+4$. In these cases, the principal symbol of the equations would vanish, making the strong-coupling issue arise again. 

If we allow $N\to \infty$ there are infinitely many choices of the operator $\mathcal{F}$ that, besides avoiding strong coupling pathologies, actually yield ghost-free non-local dynamics around maximally symmetric backgrounds.
A particularly interesting and natural choice is
\begin{equation}\label{Fnonlocal}
\mathcal{F}=\frac{e^{\Omega(\hat{\mathcal{D}})}-\mathbb{I}}{\hat{\mathcal{D}}}\, ,
\end{equation}
where $\Omega(z)$ is an entire function with $\Omega(0)=0$ and $\hat{\mathcal{D}}$ is the operator obtained from linearizing \eqref{Ehat} around an arbitrary background metric, $g_{ab}\to g_{ab}+h_{ab}$, 
\begin{equation}\label{Dhatdef}
    \delta \hat\E_{ab}(h)=\hat{\mathcal{D}}_{ab}{}{}^{cd}h_{cd}\, .
\end{equation}
If the background is flat, this reduces to the linearized Einstein tensor and $\hat{\mathcal{D}}_{ab}{}{}^{cd}$ is the Lichnerowicz operator. In a general background it will no longer coincide with the Lichnerowicz operator but, importantly, it is always a second-order operator.

In order to show that this choice leads to ghost-free dynamics, let us study linear perturbations of \req{NLQTv2} around a solution satisfying $\hat{\E}_{ab}=0$. This includes perturbations around arbitrary maximally symmetric vacua, but also around arbitrary spherically symmetric vacuum spaces. 
The expansion of \req{NLQTv2} to quadratic order in the metric perturbation $h_{ab}$ yields 
\begin{equation}\label{NLQTlinear1}
\begin{aligned}
S_{\rm NLQT}^{(2)}=-\int \frac{\mathrm{d}^Dx \sqrt{|g|} }{16\pi G}\bigg[\frac{1}{2}h^{ab} \delta\E_{ab}(h)\\
+\frac{1}{2}\delta\hat{\E}^{ab}(h) \mathcal{F}_{ab}{}{}^{cd} \delta\hat{\E}_{cd}(h)\bigg]\, ,
\end{aligned}
\end{equation}
where $
    \delta\E_{ab}(h)=\mathcal{D}_{ab}{}{}^{cd}h_{cd}
$ represents the linearized QT equations of motion \eqref{eomQT}
and, in general, $\mathcal{D}_{ab}{}{}^{cd}\neq \hat{\mathcal{D}}_{ab}{}{}^{cd}$, except in several cases  that we consider below. Here, everything except $h_{ab}$ is a background quantity but we do not introduce additional notation to reduce the clutter. 

Variation of \eqref{NLQTlinear1} with respect to $h_{ab}$ yields the linearized equations, but the computation becomes clearer if we promote $\delta\hat{\E}^{ab}(h)$ to be an independent variable $e_{ab}$ by introducing a Lagrange multiplier $\lambda^{ab}$, 
 \begin{equation}\label{NLQTlinear2}
\begin{aligned}
\delta\hat{\E}^{ab} \mathcal{F}_{ab}{}{}^{cd} \delta\hat{\E}_{cd} \rightarrow 
e^{ab} \mathcal{F}_{ab}{}{}^{cd} e_{cd}+2\lambda^{ab}(e_{ab}-\delta\hat{\E}_{ab})\, .
\end{aligned}
\end{equation}
Varying the action with respect to $\lambda^{ab}$ yields
\begin{equation}\label{habeq}
e_{ab}=\delta\hat{\E}_{ab}(h)\, ,
\end{equation}
and introducing it back one would recover the original action, showing that both are equivalent. Variation with respect to $h_{ab}$ then yields
\begin{equation}\label{lambdaabEq}
\delta\E_{ab}(h)-\delta\hat{\E}_{ab}^{\dagger}(\lambda)=0\, ,
\end{equation}
where $\delta\hat{\E}_{ab}^{\dagger}$ represents the adjoint operator of $\delta\hat{\E}_{ab}$ under the scalar product defined by integration over spacetime. We observe that, while $\delta\E_{ab}$ is always self-adjoint because it comes from the Euler-Lagrange variation of an action, $\delta\hat{\E}_{ab}$  is not in general. 
Finally, the variation with respect to $e_{ab}$ gives the equation
\begin{equation}\label{eabEq}
\lambda_{ab}+\mathcal{F}^{+}_{ab}{}{}^{cd} e_{cd}=0\, . 
\end{equation}
where 
$
    \mathcal{F}^{+}_{ab}{}{}^{cd}\equiv \frac{1}{2}(\mathcal{F}_{ab}{}{}^{cd} +\mathcal{F}^{\dagger}_{ab}{}{}^{cd})
$
is the self-adjoint part of $\mathcal{F}$. 

Now let us restrict ourselves to perturbations for which $\nabla^c \nabla^d P_{acbd}=0$ is still identically satisfied. This is the case for general perturbations around maximally symmetric spacetimes and also for SS perturbations of SS spacetimes. If the perturbations satisfy $\nabla^c \nabla^d P_{acbd}=0$, then one has $\delta\E_{ab}(h)=\delta\hat{\E}_{ab}(h)=\delta\hat{\E}_{ab}^{\dagger}(h)$ and the operator $\hat\D$ becomes self-adjoint when restricted to those backgrounds or to the corresponding subspace of perturbations. Applying $\delta\hat{\E}_{ab}$ on \eqref{eabEq} and using \eqref{habeq} and \eqref{lambdaabEq}, we get
$e_{ab}+\delta\hat{\E}_{ab}\left(\mathcal{F}^{+} e\right)=0$.  
Using \eqref{Dhatdef}, this can be written explicitly as
\begin{equation}\label{eabEq2}
\left(\delta_{(a}{}^{c}\delta_{b)}{}^{d}+\hat{\mathcal{D}}_{ab}{}{}^{ef}\mathcal{F}^{+}_{ef}{}{}^{cd}\right)e_{cd}=0\, . 
\end{equation}
Finally, if we choose $\mathcal{F}$ as in \req{Fnonlocal}, we observe that $\mathcal{F}^{+}=\mathcal{F}$ in the cases we are considering (because it is built from $\hat{\mathcal{D}}$, that becomes self-adjoint), so the equation \req{eabEq2} becomes
\begin{equation}\label{eabEq3}
e^{\Omega\left(\hat{\mathcal{D}}_{ab}{}{}^{cd}\right)}e_{cd}=0\, . 
\end{equation}
Since $\Omega$ is an entire function, the operator $e^{\Omega\left(\hat{\mathcal{D}}\right)}$ has zero kernel, and therefore the only solution of this equation is $e_{ab}=0$. This further implies that the only solutions to the full linearized equations are those of the original QT theory, $\delta \hat{\E}_{ab}(h)=0$.

{\bf Linearized equations and graviton propagator.}
In the case of maximally symmetric backgrounds, the above result means that the only propagating mode is the massless graviton, as the linearized equations of QT gravity coincide with the Einstein ones. However, the crucial difference is that now there are no extra strongly coupled modes, since there is no order reduction in the principal symbol of the equations  and all the higher-derivative terms get combined into a zero-kernel form factor that does not introduce new degrees of freedom. As a result, the spectrum remains automatically and genuinely ghost-free around maximally symmetric vacua.

To see this in the case of flat spacetime, note that $\hat{\E}_{ab}=G_{ab}+\ldots$, where the ellipsis denote terms that are at least quadratic in the curvature. 
Since $\hat{\mathcal{D}}_{ab}{}{}^{cd}$ corresponds in this case to the Lichnerowicz operator, the equivalent quadratic action  $S_{\rm NLQT}^{(2)}$ reads 
\begin{align}
\notag
&\hspace{-0.25cm}S_{\rm NLQT}^{(2)}=\int \frac{\mathrm{d}^Dx \sqrt{|g|} }{16\pi G}\Big[R+G_{ab} \left(\frac{E_{\rm t}(\Box)-1}{\Box} \right)G^{ab}\\
& \hspace{-0.25cm}+\frac{(D-2)^2}{4(D-1)}R\left ( \frac{1- E_{\rm s}(\Box)}{(D-2)\Box}-\frac{E_{\rm t}(\Box)-1}{\Box}\right)R\Big]\, ,
\label{eq:quadactflat}
\end{align}
and we defined
\begin{align}
  E_{\rm t}(\Box)\equiv e^{\Omega(-\Box/2)}\,, \quad E_{\rm s}(\Box)\equiv e^{\Omega((D-2)\Box/2)}\,,
  \label{eq:deffunctions}
\end{align}
where $\Box$ stands for the D'Alembertian in flat space $\eta_{ab}$. Let us couple the gravitational perturbations to some stress energy tensor $T_{ab}$:
\begin{equation}
\delta \mathcal{E}^{\rm NLQT}_{ab}=8 \pi G T_{ab}    \,,
\label{eq:eqmatt}
\end{equation}
where $\delta \mathcal{E}^{\rm NLQT}_{ab}$ corresponds to the linearized equations obtained from \eqref{eq:quadactflat}. Define $\bar{h}_{ab}=h_{ab}-\frac{1}{2} h \, \eta_{ab}$ and impose the de Donder gauge $\nabla^a \bar{h}_{ab}=0$. The solution to \eqref{eq:eqmatt} in momentum space, $\nabla_a \rightarrow i k_a$, reads
\begin{equation}
\bar{h}_{ab}=\mathcal{G}_{ab,cd}(k) T^{cd}\, ,
\end{equation}
where the propagator $\mathcal{G}_{ab,cd}(k)$ is given by
\begin{equation}
\label{eq:propanlqt}
\frac{\mathcal{G}_{ab,cd}(k)}{16\pi G}= \frac{\theta_{a\langle c}\theta_{d\rangle b}}{k^2E_{\rm t}(-k^2)}+\frac{\theta_{ab}\theta_{cd}}{(D-1) k^2 E_{\rm s}(-k^2)}\, .
\end{equation}
Here, $\theta_{ab}\equiv \eta_{ab}-\frac{k_{a}k_{b}}{k^2}$ and the angle brackets stand for the symmetric and traceless part. The physical degrees of freedom are identified by examining the poles of the propagator. As the functions $E_{\rm t}$ and $E_{\rm s}$ in \eqref{eq:deffunctions} have no zeroes anywhere in the complex plane, the pole structure of \eqref{eq:propanlqt} is exactly the same as in GR, with a single pole at $k^2=0$. Therefore, the theory only propagates a unitary massless graviton.

Besides the absence of new degrees of freedom, the theory becomes potentially super-renormalizable if the propagator decays in the UV faster than $|k^2|^{-D/2}$ \cite{Biswas:2005qr,Biswas:2011ar,Modesto:2011kw,Tomboulis:1997gg}. Since the argument of the entire function $\Omega$ in the form factors $E_{\rm t}$ and $E_{\rm s}$ has opposite sign, we demand that $\Omega(x)\to +\infty$ for both $x\to \pm \infty$, so that both form factors grow unbounded in the UV. The simplest choice is 
\begin{equation}
\label{eq:Omegafacil}
    \Omega(x)= \gamma^2 x^2\, , \quad \gamma >0\,,
\end{equation}
leading to an exponentially suppressed propagator. 

{\bf Regular black holes from pure gravity.}
Our results also apply to the case of SS perturbations of SS spacetimes, and they imply a perturbative form of Birkhoff's theorem. To see this, let us consider a SS solution of QT gravity which, by construction, is also an exact solution of its nonlocal version. As we have seen, in standard QT gravity any such solution is necessarily static on account of Birkhoff's theorem and it is uniquely determined by the total mass of the spacetime---{\emph{e.g.}, the Hayward black hole in \req{hayw}. 

A natural question is whether these solutions remain the unique SS vacuum solutions of the nonlocal completion \eqref{NLQTv2}. Our results imply that, at the very least, these solutions do not admit any non-trivial continuous deformation. In fact, if the form factor is chosen as in \eqref{Fnonlocal} and we consider an arbitrary SS deformation of the solution, then this deformation must necessarily satisfy the original linearized equations of QT gravity. However, we already know that the (full) SS equations of QT gravity imply unicity of solutions. Therefore, we conclude that spherical waves do not exist and the only possible effect of this perturbation is to shift the mass of the solution by a constant term. This provides a linearized form of Birkhoff's theorem in NLQT gravities. This also means that, although the equations of motion now contain an infinite number of derivatives, the dynamics of the SS sector remains healthy.  

These results provide, to our knowledge, the first examples of nonlocal gravity theories admitting exact RBH solutions. Remarkably, the singularity resolution mechanism in these vacuum solutions is fully governed by the limiting curvature induced by the infinite tower of nonlinear curvature terms in the action, rather than by nonlocal effects. 


{\bf Regular black holes and matter sources.}
Nonlocal extensions of GR have frequently been argued to provide a mechanism for regulating spacetime singularities \cite{Buoninfante:2022ild}. Unfortunately, the highly intricate structure of their equations usually restricts analyses to perturbative regimes and approximate methods \cite{Frolov:2015bta,Frolov:2015bia,Boos:2018bxf,Buoninfante:2018stt,Buoninfante:2018xiw,Burzilla:2020utr,Modesto:2011kw,Zhang:2014bea,Giacchini:2018gxp}. 
Consequently, standard arguments mainly demonstrate the smearing effect of the nonlocal operators on point-like singular sources in the weak-field approximation. Nevertheless, at the same time, these theories typically admit all exact singular vacuum solutions of GR \cite{Li:2015bqa,Calcagni:2017sov}.

The contrast with local QT gravities is notable. Indeed, despite admitting RBHs as the unique SS solutions, the Newtonian potential of local QT gravities is identical to the Einstein gravity one and, therefore, divergent at $r=0$---although this is cured in the nonlinear regime. Additionally, certain minimally coupled singular-matter sources have been shown to give rise to singular geometries \cite{Bueno:2025tli,Bueno:2026dln}. 

NLQT gravity captures both types of regulating effects. Indeed, consider a point mass with $T_{00}= m \delta^{D-1}(\vec{r})$. The metric generated by this source can be written as
\begin{equation}
   \hspace{-0.1cm} \mathrm{d}s^2=-\left[1+2 \Phi_1(r)\right] \mathrm{d}t^2+ \left[1-2 \Phi_2(r)\right] \mathrm{d}\vec{x}_{D-1}^2\,,
\end{equation}
where $r$ is the radial coordinate in Euclidean space $\mathrm{d}\vec{x}_{D-1}^2$. For the theory \eqref{eq:Omegafacil}, the Newtonian potentials $\Phi_1(r)$ and $\Phi_2(r)$ behave near $r=0$  as
\begin{equation}
\Phi_i=\frac{Gm}{\gamma^{(D-3)/2}}\left [\mathfrak{a}_i^{(0)}+ \mathfrak{a}_i^{(2)}\frac{r^2}{\gamma} +\mathcal{O}\left (\frac{r^4}{\gamma^2} \right)\right]\,,
\end{equation}
where $\mathfrak{a}_i^{(0)}$ and $\mathfrak{a}_i^{(2)}$ are some constants that are given in the appendix. Therefore, the Newtonian potentials and the geometry are regular around $r=0$. As $r \rightarrow \infty$,  one recovers the usual $r^{3-D}$ falloff and $\Phi_1 \simeq (D-3)\Phi_2$. 
If the mass $m$ is small enough, the linearized solution remains a good approximate solution of the full theory for all values of $r$. 

To sum up, all exact vacuum SS solutions continuously connected with the GR ones are regular due to the limiting-curvature effect of QT gravity, and weak fields from point sources are regular due to nonlocal smearing of the source. 
Finding exact SS solutions in the presence of matter is challenging, but we expect all of them to remain regular from a combination of both mechanisms. 
For instance, a delta source inside the inner core of a vacuum RBH would effectively become Gaussian-like and potentially modify the metric in the region $r \lesssim \sqrt{\gamma}$. If we choose our action scales such that $\gamma\ll \alpha$, this represents a tiny portion of the inner core. 
On the other hand, if we choose $\gamma\sim \alpha$, the effects of the nonlocal smearing could extend up to the inner horizon scale.



{\bf Conclusions.}
We have introduced a new class of nonlocal theories of gravity that preserve the pure-gravity RBH solutions of QT theories featuring infinite towers of corrections to the Einstein–Hilbert action. Simultaneously, their nonlocal character removes the strong-coupling instabilities, ensures that the theories are ghost-free and renders them potentially super-renormalizable around flat space.

NLQT theories combine two distinct mechanisms for resolving spacetime singularities: first, a limit on how curved spacetime can be, arising from curvature nonlinearities in the action---analogous to the limiting speed imposed by special relativity~\cite{UsLong}; and second, the nonlocal smearing of matter sources. It would be interesting to further study the interplay between matter and spacetime regularity within these theories. In particular, the collapse of SS matter perturbations around a vacuum RBH may prove computationally tractable. 


In the search for a theory of quantum gravity, it has become increasingly clear that infinite towers of higher-derivative corrections are unavoidable. However nature circumvents singularity formation, it is plausible that, at least at an effective level, the relevant physics involves a combination of the two mechanisms brought together here for the first time.


\emph{Acknowledgments.}  PB was supported by a Proyecto de Consolidación Investigadora (CNS 2023-143822) from Spain’s Ministry of Science, Innovation and Universities, and by the grant PID2022-136224NB-C22, funded by MCIN/AEI/10.13039/501100011033/FEDER, UE. PAC was supported by a Ramón y Cajal fellowship (RYC2023-044375-I) and by a Proyecto de Generación de Conocimiento
(PID2024-155685NB-C22) from Spain’s Ministry of Science, Innovation and Universities, and by an ``Attract RyC'' grant (Agujeros negros, f\'isica fundamental y ondas gravitacionales) from the University of Murcia. RAH is supported by a Willmore Fellowship at Durham University. {\'A}JM was supported by a Juan de la Cierva contract (JDC2023-050770-I) from Spain’s Ministry of Science, Innovation and Universities.

\newpage


\onecolumngrid  
\begin{center}  
{\Large\bf Appendix} 
\end{center} 
\appendix 

\section{Linearized equations and Newton potential}

Let us start from the quadratic action controlling perturbations around flat spacetime: 
\begin{align}
\hspace{-0.25cm}S_{\rm NLQT}^{(2)}=\int \frac{\mathrm{d}^Dx \sqrt{|g|} }{16\pi G}\Big[R+G_{ab} \left(\frac{E_{\rm t}(\Box)-1}{\Box} \right)G^{ab}+\frac{(D-2)^2}{4(D-1)}R\left ( \frac{1- E_{\rm s}(\Box)}{(D-2)\Box}-\frac{E_{\rm t}(\Box)-1}{\Box}\right)R\Big]\, .
\label{eq:quadactflatapp}
\end{align}
where
\begin{align}
  E_{\rm t}(\Box)\equiv e^{\Omega(-\Box/2)}\,, \quad E_{\rm s}(\Box)\equiv e^{\Omega((D-2)\Box/2)}\,,
  \label{eq:deffunctionsapp}
\end{align}
for a certain entire function $\Omega$. Note that $\Box$ stands for the D'Alembertian in flat spacetime $\eta_{ab}$. The linearized equations $\delta \mathcal{E}_{ab}$ read:
\begin{align}
\label{eq:minkpertv2app}
\delta  \mathcal{E}_{ab}= E_{\rm t}(\Box)\, G_{ab}^{(1)}- \frac{D-2}{2(D-1)}\left(\frac{ E_{\rm t}(\Box)- E_{\rm s}(\Box)}{ \Box}\right)\left(\nabla_{a}\nabla_{b}-\eta_{ab}\Box\right)R^{(1)} \, ,
\end{align}
where and $G_{ab}^{(1)}$ and $R^{(1)}$ stand for the linearized Einstein tensor and Ricci scalar, respectively. Now, let us consider scalar perturbations around Minkowski spacetime in the Newtonian (conformal) gauge:
\begin{equation}
    \mathrm{d}s^2=-\left(1+2 \Phi_1\right) \mathrm{d}t^2+ \left(1-2 \Phi_2\right) \mathrm{d}\vec{x}_{D-1}^2\,,
\end{equation}
where  the potentials $\Phi_1$ and $\Phi_2$ are assumed to depend solely on the radial coordinate $r$ associated with the $(D-1)$-dimensional Euclidean space. Consider the potentials to be sourced by a point particle of mass $m$, whose stress-energy tensor is given by  $T_{ab}=\rho u_a u_b$ with $u_a=\delta_a^t$ and $\rho=m \delta^{D-1} (\vec{r})$. Using \eqref{eq:minkpertv2app}, the whole set of linearized equations reduces to the resolution of the traced equations and the $tt$ component, respectively given by
\begin{align}
    \Box_r E_{\rm s}(\Box_r) (\Phi_1-(D-2)\Phi_2)&=-\frac{8 \pi G m}{(D-2)}\delta^{D-1} (\vec{r})\,, \\ 
   \frac{E_{\rm t}(\Box_r)-E_{\rm s}(\Box_r)}{(D-1)} \Box_r  (\Phi_1-(D-2)\Phi_2)+E_{\rm t}(\Box_r) \Box_r \Phi_2 &=\frac{8 \pi G m}{(D-2)} \delta^{D-1} (\vec{r})\,,
   \label{eq:newtoneqsapp}
\end{align}
where $\Box_r F=\frac{1}{r^{D-2}} \frac{\mathrm{d}}{\mathrm{d}r} \left (r^{D-2}\frac{\mathrm{d}F}{\mathrm{d}r}  \right)$ for any function $F$ depending on $r$. Going into $(D-1)$-dimensional Fourier space, so that $\Box_r \rightarrow -\mathbf{k}^2$ and $\delta^{D-1} (\vec{r}) \rightarrow (2\pi)^{(1-D)/2}$, one may write \eqref{eq:newtoneqsapp} in Fourier space and solve for the Fourier-transformed potentials $\Phi_{1,\mathbf{k}}$ and $\Phi_{2,\mathbf{k}}$, which take the form
\begin{equation}
   \begin{pmatrix}
    \Phi_{1,\mathbf{k}} \\ \Phi_{2,\mathbf{k}}
    \end{pmatrix} =\frac{8 \pi G m }{(2\pi)^{(D-1)/2}(D-1) (D-2) \mathbf{k}^2} \begin{pmatrix}
    e^{-\Omega\left (- \frac{(D-2)\mathbf{k}^2}{2} \right)}-(D-2)^2e^{-\Omega \left (\frac{\mathbf{k}^2}{2} \right)} \\- e^{-\Omega\left (- \frac{(D-2)\mathbf{k}^2}{2} \right)} -(D-2) e^{-\Omega \left (\frac{\mathbf{k}^2}{2} \right)}
    \end{pmatrix}\,.
\end{equation}
The potentials in real space are obtained through an inverse Fourier transform:
\begin{equation}
    \begin{pmatrix}
    \Phi_1 \\ \Phi_2
    \end{pmatrix}= r^{\frac{3-D}{2}} \int_0^\infty \mathrm{d}\mathbf{k} \, \mathbf{k}^{\frac{D-1}{2}} J_{\frac{D-3}{2}}(\mathbf{k}r) \begin{pmatrix}
    \Phi_{1,\mathbf{k}} \\ \Phi_{2,\mathbf{k}}
    \end{pmatrix}\,,
    \label{eq:newpotsapp}
\end{equation}
where $J_\nu(x)$ stands for the Bessel function of the first kind with index $\nu$. Observe that ensuring the asymptotic behavior $\Omega(\vert x \vert\rightarrow  \infty) \sim  x^{2n}$ for $n \in \mathbb{N}$ guarantees the convergence of the integrals \eqref{eq:newpotsapp}.

The simplest non-trivial choice of entire function corresponds to $\Omega(x)=\gamma^2 x^2$ with $\gamma >0$. In this case, the potentials $\Phi_1$ and $\Phi_2$ may be computed analytically and one gets:
\begin{align}
\label{eq:potnewapp}
    \Phi_1(r)&=\frac{Gm}{ (2\pi \gamma)^{(D-3)/2}(D-1)} \left[ (D-2)^{\frac{1-D}{2}} \mathcal{W}\left (-\frac{r^2}{2(D-2)\gamma} \right) -(D-2) \mathcal{W}\left (-\frac{r^2}{2\gamma} \right)\right] \, ,\\
    \label{eq:potnewapppsi}
    \Phi_2(r)&=-\frac{G m}{ (2\pi \gamma)^{(D-3)/2}(D-1) } \left[ (D-2)^{\frac{1-D}{2}} \mathcal{W}\left (-\frac{r^2}{2(D-2)\gamma} \right) +\mathcal{W}\left (-\frac{r^2}{2\gamma} \right)\right]\,,
\end{align}
where we have defined
\begin{align}
\notag
    \mathcal{W}\left (z \right)&\equiv 2^{\frac{3-D}{2}}  \pi  \Gamma \left(\frac{D-3}{4}\right) \, _1\tilde{F}_3\left(\frac{D-3}{4};\frac{1}{2},\frac{D-1}{4},\frac{D+1}{4};\frac{z^2}{16}\right)\\&+2^{-\frac{1}{2} (D+1)} \pi   z \Gamma \left(\frac{D-1}{4}\right) \, _1\tilde{F}_3\left(\frac{D-1}{4};\frac{3}{2},\frac{D+1}{4},\frac{D+3}{4};\frac{z^2}{16}\right)\,,
\end{align}
where ${}_1\tilde{F}_3(z)$ stands for the regularized generalized hypergeometric function. Expanding the result for the Newtonian potentials \eqref{eq:potnewapp} and \eqref{eq:potnewapppsi} around $r=0$, one gets
\begin{equation}
    \Phi_i=\frac{Gm}{\gamma^{(D-3)/2}}\left [\mathfrak{a}_i^{(0)}+ \mathfrak{a}_i^{(2)}\frac{r^2}{\gamma} +\mathcal{O}\left (\frac{r^4}{\gamma^2} \right)\right]\,,
\end{equation}
where
\begin{equation}
\mathfrak{a}_i^{(0)}\equiv -\frac{ \Gamma\left (\frac{D-3}{4}\right)  \left ( (D-2)^{i-\frac{D+3}{2}}+(-1)^{-i} \right)}{(2\pi )^{(D-3)/2}  \Gamma\left (\frac{D-1}{2}\right )(D-1) (2-D)^{i-2}}\,, \quad \mathfrak{a}_i^{(2)}\equiv \pi \frac{ \Gamma\left (\frac{D-1}{4}\right)  \left ( (D-2)^{i-\frac{D+5}{2}}+(-1)^{-i} \right)}{(2\pi )^{(D-1)/2} \Gamma\left (\frac{D+1}{2}\right )(D-1) (2-D)^{i-2}} \,.
\end{equation}
On the other hand, both potentials $\Phi_1$ and $\Phi_2$ decay asymptotically as $r^{3-D}$, obtaining $\Phi_1 \simeq (D-3)\Phi_2$ in this limit---just like in GR.

\twocolumngrid

\bibliographystyle{JHEP-2}

\bibliography{Gravities}
\noindent 


\end{document}